\documentclass[11pt,onecolumn, thmsa]{article}

\usepackage{graphicx}
\usepackage{latexsym}
\usepackage{amssymb}
\usepackage{amsmath}
\usepackage{amsfonts}
\usepackage{amstext}
\usepackage{euscript}
\usepackage{epsfig}
\begin{document}

\title{Functional dissipation microarrays for classification}
\date{}
\author{D. Napoletani \thanks{ Center for Applied Proteomics and Molecular Medicine, George Mason University, Manassas, VA
20110, email: dnapolet@gmu.edu}, D. C. Struppa \thanks{ Department
of Mathematics and Computer Science, Chapman University, Orange,
CA 92866}, T. Sauer \thanks{ Department of Mathematical Sciences,
George Mason
University, Fairfax, VA 22030}\\
and\\
V. Morozov$^\S$, N. Vsevolodov$^\S$, C. Bailey \thanks{ National
Center for Biodefense and Infectious Diseases, George Mason
University 20110, Manassas, VA }}

\maketitle

\begin{abstract}

In this article, we describe a new method of extracting
information from signals, called functional dissipation, that
proves to be very effective for enhancing classification of high
resolution, texture-rich data. Our algorithm bypasses to some
extent the need to have very specialized feature extraction
techniques, and can potentially be used as an intermediate,
feature enhancement step in any classification scheme.

Functional dissipation is based on signal transforms, but uses the
transforms recursively to uncover new features. We generate a
variety of masking functions and `extract' features with several
generalized matching pursuit iterations. In each iteration, the
recursive process modifies several coefficients of the transformed
signal with the largest absolute values according to the specific
masking function; in this way the greedy pursuit is turned into a
slow, controlled, dissipation of the structure of the signal that,
for some masking functions, enhances separation among classes.

Our case study in this paper is the classification of
crystallization patterns of amino acids solutions affected by the
addition of small quantities of proteins.
\\
\\
{\it Keywords:} Features enhancement, matching pursuit,
classification, non-linear iterative maps, wavelet transforms.

\end{abstract}

\clearpage

\newcommand{\C}{I\!\!\!\! C}
\def\RR{I\!\! R}
\renewcommand{\L}{L\!\! L}
\newcommand{\N}{I\!\! N}
\newcommand{\R}{I\!\! R}
\newcommand{\iP}{I\!\! P}
\newcommand{\Z}{Z\!\!\! Z}
\newcommand{\1}{1\!\!\! 1}
\newtheorem{definition}{Definition}[section]
\newtheorem{theorem}{Theorem}[section]
\newtheorem{prop}{Proposition}[section]
\newtheorem{post}{Postulate}[section]
\newtheorem{cor}{Corollary}[section]
\newtheorem{lemma}{Lemma}[section]

\section{Introduction}

In this paper we introduce the {\it functional dissipation
microarray}, a feature enhancement algorithm that is directly
inspired by experimental microarray techniques [1]. Our method
shows how ideas from biological methodologies can be successfully
turned into functional data analysis tools.

The idea behind the use of microarrays is that if a large and
diverse enough data set can be collected on a phenomenon, it is
often possible to answer many questions, even when no specific
interpretation for the data is known. The algorithm we describe
here seems particularly suitable for high resolution, texture-rich
data, and bypasses to some extent the need to preprocess with
specialized feature extraction algorithms. Moreover, it can
potentially be used as an intermediate feature enhancement step in
any classification scheme.

Our algorithm is based on an unconventional use of matching
pursuit ([2], Chapter 9; [3]). More precisely, we generate random
masking functions and `select' features with several generalized
matching pursuit iterations. In each iteration, the recursive
process modifies several of the largest coefficients of the
transformed signal according to the masking function. In this way
the matching pursuit becomes a slow, controlled dissipation of the
structure of the signal; we call this process {\it functional
dissipation}. The idea is that some unknown statistical feature of
the original signal many be detected in the dissipation process at
least for some of the random maskings.

This process is striking in that, individually, each feature
extraction with masking becomes unintelligible because of the
added randomness and dissipation, and only a string of such
feature extractions can be `blindly' used to some effect. There is
some similarity in spirit between our approach and the beautiful
results on random projections reconstructions described in [4] and
[5], with the important difference that we use {\it several}
distinct randomization and dissipation processes to our benefit so
that there is a strong non-linear dynamics emphasis in our work.
Moreover, we bypass altogether reconstruction issues, focusing
directly on classification in the original representation space.
Our method can be seen also as a new instance of ensemble
classifiers (like boosting [6] and bagging [7]), in that several
functional dissipations are generally pulled together to achieve
improvement of classification results.

Other applications of matching pursuit to classification problems
include kernel matching pursuit methods [8], and also boosting
methods, that can be interpreted essentially as greedy matching
pursuit methods where the choice of `best' coefficient at each
iteration is made with respect to more sophisticated loss
functions than in the standard case (see [9], [10], [11] chapter
10). We stress again that, while our approach uses the {\it
structure} of matching pursuits, only their rich dynamical
properties (highlighted for example in [3]) are used for the
generation of features, since in our method the whole iterative
process of modifying coefficients becomes simply {\it an instance
of non-linear iterative maps}, disjoined from approximation
purposes.

The case study of this paper is the classification of
crystallization patterns of amino acids solutions affected by
addition of small quantities of proteins; such crystallization
patterns show varied and interesting textures. The goal is to
recognize whether an unknown solution contains one of several
proteins in a database. Crystallization patterns may be
significantly affected by laboratory conditions, such as
temperature and humidity, so the degree of similarity of patterns
belonging to a same protein is subject to some variations, adding
difficulty to the classification problem. The richness of the
textures and their variability gives this classification problem
an appeal that goes beyond the specific application from which it
arises.

Our basic approach is to derive long feature vectors for each
amino acid reporter  by our generalized recursive greedy matching
pursuit. Since the crystallization patterns are generated by
stochastic processes, there is a great local variability in each
droplet and any significant feature must encode a statistical
information about the image to be part of a robust classifier, see
[10]. We can see this case study as an instance of texture
classification, and it is well established that wavelets (see [2]
chapter 5, [13]) and moments of wavelet coefficients (see [14], or
[15] for the specific problem of iris recognition) can be very
useful for these types of problems. Therefore we implemented our
method in the context of a wavelet image transform for this case
study, even though we then summarized the information obtained
with the functional dissipation with moments of the dissipated
images in the original representation space. Other choices of
statistical quantities are possible in principle according to the
specific application, since, as we show in section 2, the method
can be formulated in a quite general setting.

In section 2 we introduce the features enhancement method,
functional dissipation, to explore the feature space. In sections
3 and 4 we apply the method to the crystallization data to show
how the application of a sufficiently large number of different
functional dissipations can greatly improve the classification
error rates.

\section{Functional dissipation for classification}

In this section we introduce a classification algorithm that is
designed for cases where feature identification is complicated or
difficult. We first outline the major steps of the algorithm, and
then discuss specific implementations. In the following sections
we apply one possible implementation of this method to droplet
classification, and describe the results.

We begin with input data divided into a training set and a test
set. We then follow four steps as follows:
\begin{enumerate}
\item[A] Choose a classifier and a figure of merit that quantifies
the classification quality.

\item[B] Choose a basic method of generating features from the
input data, and then enhance the features by a recursive process
of structure dissipation, described below.

\item[C] Compute the figure of merit from A for all features
derived in B. For a fixed integer $p$, search the feature space
defined in B for the $p$ features which maximize the figure of
merit in A on the training set.

\item[D] Apply the classifier from A, using the optimal $p$
features from C, to classify the test set.
\end{enumerate}

In step A, for example, multivariate linear discrimination can be
used as a classifier. This method comes with a built-in figure of
merit, the ratio of the between-group variance to the within-group
variance. More sophisticated classifiers often have closely
associated evaluation parameters. Cross-validation or
leave-one-out error rates can be used.

Step B is the heart of the functional dissipation algorithm. The
features used will depend on the problem. For two-dimensional
images, orthogonal or over-complete image transforms can be used.
The method of functional dissipation is a way to leverage the
extraction of general purpose features to generate features with
increased classification power. This method uses the transforms
recursively to gradually modify the feature set.

Consider a single input datum $X$ and several invertible
transforms $T_k$, $k=1, \ldots, K$, that can be applied to $X$ (in
the case study to be shown later, $X$ represents a $256\times 256$
gray scale image and $T_k$ represent Discrete Wavelet Transforms).
At each iteration we select several coefficients from the input
datum $X$ and we use the mask to modify the coefficients
themselves. Fix positive integers $K,M$ and set $X_0=X$ Let $A(x)$
be a discrete valued function defined on $\Z$, which we call a
{\bf mask} or a {\bf masking function}. Apply the following {\bf
functional dissipation} steps (B1)-(B3) $K$ times. For $k=1,
\ldots, K$:

{\bf (B1):} {\it Compute the transform $T_k X_{k-1}$}

{\bf (B2):} {\it Choose a subset $\mathcal S$ of $T_k X_{k-1}$ and
collect the $M$ coefficients $C(m)$, $m=1,...,M$ in $\mathcal S$
with largest absolute value in a suitable subset.}

{\bf (B3):} {\it Apply the mask: Set $C'(m)=A(m)C(m)$, and modify
the corresponding coefficients of $T_k X_{k-1}$ in the same
fashion. Set $X_{k}=T_k^{-1}(T_k X_{k-1})'$ to be the inverse of
the modified coefficients $(T_k X_{k-1})'$}.

At the conclusion of the $K$ steps, features are generated by
computing statistics that describe the probability distribution of
$X_k$, $k=0, \ldots, K$. For example, one could use $m(h)$,
$h=3,4$, the third and fourth moments of the set (or even more
moments for large images). These statistics are used as features,
delivered by the means of functional dissipation. If we carry out
these steps for $N$ different masks $A_n$, we obtain a
$2NK+2$-dimensional feature vector for each data input (where we
counted the moments of the input images only once).

One way to view our approach is as a matching pursuit strategy
(see [2] chapter 9), but used in a new and unusual way. In
general, matching pursuit is used to find good suboptimal
approximations to a signal. The way this is done is by expanding a
function $f$ in some dictionary $\mathcal D=\{g_1,...,g_P\}$ and
by choosing the element in the dictionary $g_k$ for which
$|<f,g_k>|$ is maximum. Given an initial approximation $\tilde
f=0$ of $f$, and an initial residue $Rf=f$, we set $\tilde
f=\tilde f+<f,g_k>g_k$ and $Rf=Rf-<f,g_k>g_k$. The process is
repeated on the residue several times to extract successively
different relevant structures from the signal.

Instead, in each iteration of our algorithm we modify several of
the largest coefficients in different regions of the transformed
signal according to the random masking functions, so that only the
non-linear interaction of the signal and the masking function is
of interest and not the approximation of the underlying signal.
The residue is perturbed more and more until, in the limit, no
structure of the original signal is visible in either $Rf$ {\it
and} $\tilde f$. This is therefore a slow, controlled dissipation
of the structure of the signal. Note that the {\it structure} of
the signal is dissipated, not its energy, in other words the
images $X_k$ are allowed in principle to increase in norm as $k
\rightarrow \infty$.

We can think of the input image $X$ as initial condition of the
iterative map defined by mask and dissipation, and the application
of several dissipation processes allows to identify those maps for
which different classes of images flow in different regions of the
feature space as defined by the output of the map. It seems likely
that, for some non-linear maps, similar initial conditions will
evolve with similar dynamics, while small differences among
classes will be enhanced by the dynamics of the map and therefore
they will be detectable as the number of iterations increase. The
key question is whether a small set of masks can generate enough
variability among the dissipative processes to find such
interesting dynamics. Our results show that, for our case study,
this is the case. According to this qualitative dynamical system
interpretation, our method can be seen along the lines of
dynamical system algorithms for data analysis such as those
developed in [16] to approach numerical linear algebra problems.

In choosing the masking functions for (B1)-(B3), the guiding idea
is to have large variability in the masks themselves so that there
is a wide range of possible dynamics in the dissipations
processes, while at the same time preserving some of the structure
of the signal from one iteration to the next (in other words we
want the signals to be dissipated, but slowly). To this extent,
the masking functions are designed to assume small values so that
the image is only slightly affected at each iteration by the
change of of a subset of its coefficients as in (B1)-(B3). To
respect these limitations, we decided to take each mask as a
realization of length $M$ of Gaussian white noise with variance
$1$, we then convolve it with a low-pass filter to increase
smoothness and finally we rescale it to have a fixed maximum
absolute value.

More precisely, let $W[m]$, $m=1,...,M$ be a realization of
Gaussian white noise of length $M$, and let $g$ be a smoothing
filter defined on all integers $\Z$ such that
$g[m]=\cos^2(\frac{\pi m}{2E}) \1_{[-E,E]}[m]$ where $\1_{[-E,E]}$
denotes the function that assumes value $1$ for $m<E$ and value
$0$ otherwise (we follow here [2] page 440). Let now $\tilde
W=W*g$ be the convolution of $W$ and $g$, where $W$ is suitably
extended periodically at the boundary. Then each mask can be
written as $A=\alpha \tilde W/\max(|\tilde W|)$, where $\alpha$ is
a small real number. The larger $E$, the smoother the mask $A$,
but the fact that each underlying $W$ is a random process assures
the necessary variability. We repeat this process $N$ times
choosing several values of $E$ and $\alpha$ to generate curves of
the type shown in Figure 1.
\begin{figure}
  \includegraphics[ angle=0 ,width=0.7\textwidth]{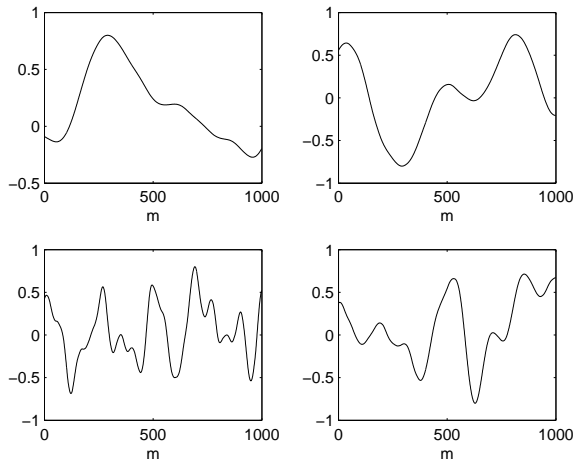}
  \caption{Examples of masking functions (defined for
coefficient indexes $m=1,...,1000$) used in the iterative process
(B1)-(B3).}
\end{figure}

{\bf Remark 1:} {\it For our case study in section 3, we choose
$E$ to be distributed on $[M/3,0]$, and we choose a logarithmic
distribution on this set so that smoother masks are favored.
Specifically we take $M=1000$ and $E$ uniformly in the set
$\mathcal E=\{299,139,64\}$. We choose $\alpha<1$ (specifically,
$\alpha=0.8$) to cause slow dissipation.}

The randomness of the specific choice of masks is used to allow a
wide spanning of possible mask shapes, and we expect more general
classes of maskings to be suitable for this method. On the other
hand the question of which is the smallest class of masks that
allows effective feature extraction is still open.\footnote{A
patent application has been filed on the method described in this
paper with U.S. Provisional Patent Application Number 60/687,868,
file date 6/7/2005.}.

\section{Case Study: Droplet Classification}

In this section we describe the case study that motivated this
research, the classification of proteins by their effect on
crystallization patterns of amino acids used as reporter
substances. We restrict ourselves to a database of four classes
:$3$ proteins and control. The proteins are albumin from chicken
egg white, hemoglobin from bovine blood and lysozyme, which we
denote respectively as $P1$, $P2$ and $P3$ respectively. These
proteins were added to solutions of one amino acid, namely
leucine, denoted by $L$. The control solution without protein will
be denoted as $Water$. Both the number of proteins to classify and
the number of reporters can increase in practice, but we restrict
our analysis to this smaller set for our purposes as already with
these few classes of proteins we can see the basic difficulty of
distinguishing even by eye some of the patterns.

A crucial advance in the experimental study of droplets has been
the ability to generate quickly a large number of droplets to
which different amino acids have been added. Remarkably, the
addition of proteins to the amino acid reporters can have very
different effects on the crystallization patterns. In some cases
there is no visible difference between droplets of amino acid plus
$Water$ (control) and droplets of amino acid plus protein; for
some other proteins the resulting crystallization patterns are
very different from control [17]. Clearly the different behavior
of solutions for different amino acids can greatly facilitate
classification.

\begin{figure}
\includegraphics[angle=0,width=1\textwidth]{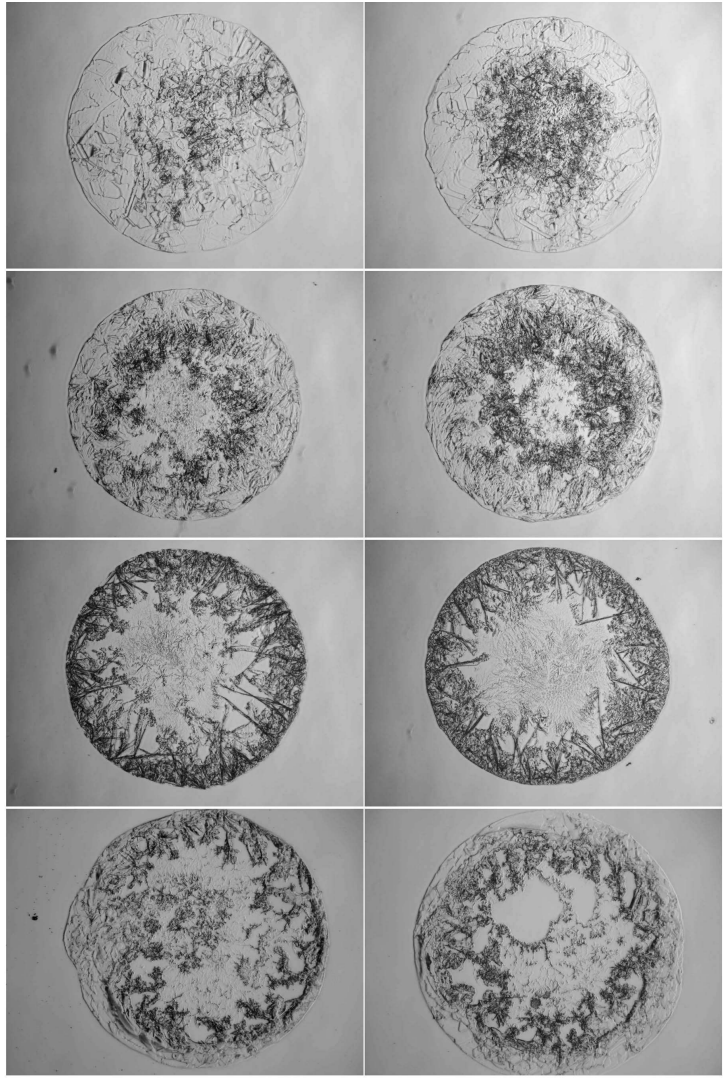}
\caption{Each row of this figure shows, from top to bottom, two
representative patterns of crystallization of amino acid $A$ to
which was added respectively: $Water$, $P1$, $P2$,$P3$.}
\end{figure}
In Figure 2, each row shows two representative patterns of the
amino acid $L$ solutions with the three proteins $P1$, $P2$, $P3$
and $Water$ added. Note in particular that the appearance of
droplets for $P1$ and $Water$ show several similarities (for
example large, glassy crystals at the boundary) and that there is
a significant degree of variability within each class.

\section{Results and Discussion}

In this section we discuss a particular implementation of the
pattern classification algorithm of Section 2, and the results of
applying it to the droplets. We used for our analysis 20
gray-scale images of crystallization patterns  for each
combination of amino acid $L$, with proteins $P1$,$P2$, $P3$ and
with $Water$. For each droplet $X_{ij}$, i.e. the $i$-th image
instance of protein $j$ with addition of amino acid $L$, we reduce
the size of the image to a matrix of $256$ by $256$ pixels and we
normalize the image so that its mean as an array is zero and its
variance is one.

For purposes of comparison, we will apply the steps (B1)-(B3) in
several ways. First we set the number of iterations of the
dissipation to zero ($K=0$), and we apply the process only once,
which corresponds to a direct evaluation of the $2$ moments of the
images. Then we apply the algorithm  choosing the $2$ best
features with several masks and no dissipation ($K=1$) and with
several masks and function dissipation ($K>1$). The masking
functions we use are the Gaussian processes described  at the end
of section 2 with parameters as in remark 1. Subsequently we use,
for the case $K=1$ and $K>1$, a larger number of features (namely
$6$, one for each distinct pair of classes), and we show that
error rates are very low in this setting when $K>1$.

In all cases we apply to the selected features a general purpose
classification algorithm such as a 3-nearest neighborhood (3-NN)
classifier to the output of (B1)-(B3) and we test each case by
dividing the $20$ instances of feature vectors for each class
randomly into a training set of $15$ instances and a test set of
$5$ instances. We train  the classifier on the training set and
then test it on the remaining $5$ instances of each class. Note
that we are interested in identifying each distinct protein and
control reporters as well, so the total number of classes that we
try to discriminate are $4$, i.e. the three proteins and $Water$.

We repeated each classification scheme $10000$ times, using
different divisions into training and testing sets to obtain
estimated of the misclassification error rates for each class.

The transforms $T_k$ in steps (B1)-(B3) are set to be the discrete
wavelet transform with Daubechies mother wavelet ($8$ vanishing
moments). The restricted subset of action $\mathcal S$ (as in
(B2)) is the detail level at scale $2^{-5}$ (highlighting
intermediate scale details of the image) for $k$ even and the
detail level at scale $2^{-7}$ (fine scale details) for $k$ odd
(see [2] section 7.7 for more on two dimensional orthogonal
wavelet bases).

{\bf Remark 2:}{\it The signal transforms need to be selected
according to the specifics of the problem. The potential of our
method is that the features exploration performed by functional
dissipation utilizes variations of general purpose signal
transforms, avoiding the need to exactly adapt the transform to
the characteristics of the classes of signals.}

{\bf Remark 3:}{\it Moments of images for large iterations can
show sometimes a very distinct order of magnitude for different
classes. While this is good in general for classification, it is
not so good for selecting suitable `best' features. The reason is
that the differences among some of the classes can be flattened if
they are very close with respect to other classes. Therefore we
take the logarithm of the absolute value of moments in all the
following classification schemes, and we scale them to have joint
variance $1$ and mean $0$.}

First, with $K=0$ no dissipation or coefficient masking occurs.
For each droplet image a two dimensional feature vector is
extracted with (B1)-(B3) by computing the log of the norm of the
third and forth moments of the images. The 3-NN classifier trained
on these features achieves classification with a low degree of
accuracy especially for $P3$: the estimated misclassification
errors for the three test proteins and $Water$ are: for $P1$
$0.062$; for $P2$ $0.050$ for $P3$ $0.227$ and for $Water$
$0.099$. This poor classification is not surprising as statistics
of images before a suitable image transform are generally
considered a weak classifier.

Second, we now take $N=100$ random functions $A_n$, $n=1,...,N$
defined on $\Z$ as in section 2. For each $A_n$ we apply steps
(B1)-(B3) with $K=1$ iterations and  we select $M=1000$
coefficients. The repeated use of (B1)-(B3) for each masking $A_n$
gives a $2NK+2=202$-dimensional feature vector for each image
droplet (we counted only one time the moments of the input
images). We select now the `best' $p$ features (say for example
$p=2$ to be consistent with the previous case) for which the ratio
of between-class variance over within-class variance for the
training sets were maximal (see [11] page 94 for more on such
notion).

If we use only these $2$ features in the 3-NN classification of
the test sets, then the estimated misclassification errors for the
four classes are: for $P1$ $0.050$; for $P2$ $0.050$ for $P3$
$0.481$ and for $Water$ $0.177$. Not only there is no improvement
over the case $K=0$, but results significantly worsen for $P3$ and
$Water$ (this is so even when we let $p$ increase). This is at
first sight puzzling as we are using also the features collected
in the case $K=0$, but the features selection may well choose a
feature that has better separation for some classes, but worse for
others. For other tests sets we did not observe such peculiar
phenomenon, but the error rates are consistently of the same order
of the case $K=0$.

Third, we take $N=100$ random masks $A_n$, $n=1,...,N$ defined on
$\Z$ as in section 2 and we now turn on the function dissipation
technique by setting $K=20$ in the steps (B1)-(B3). At each
iteration we select $M=1000$ coefficients. If we repeat (B1)-(B3)
for each masking $A_n$, we obtain a $2NK+2=4002$-dimensional
feature vector for each image droplet. We select now the `best'
$p=2$ features for which the ratio of between-class variance over
within-class variance for the training sets were maximal. If we
use only these $2$ moments in the 3-NN classification of the test
sets, then the estimated misclassification errors for the four
classes are:for $P1$ $<10^{-4}$; for $P2$ $0.051$ for $P3$ $0.059$
and for $Water$ $0.019$. For $P3$ and $Water$, this is an
excellent improvement with respect to the feature selection based
on logarithms of moments of the original images (the first case we
considered with $K=0$), while errors are of the same order for
$P1$ and $P2$.

The progression from $K=1$ to $K>1$ shows in a compelling way that
both masking function {\it and} dissipation are essential
ingredients of the method to improve error rates and masking alone
(which can be seen as a type of nonlinear projection) is not
sufficient, or may actually, in some cases, be harmful by itself
for classification.
\begin{figure}
  \includegraphics[ angle=0 ,width=0.7\textwidth]{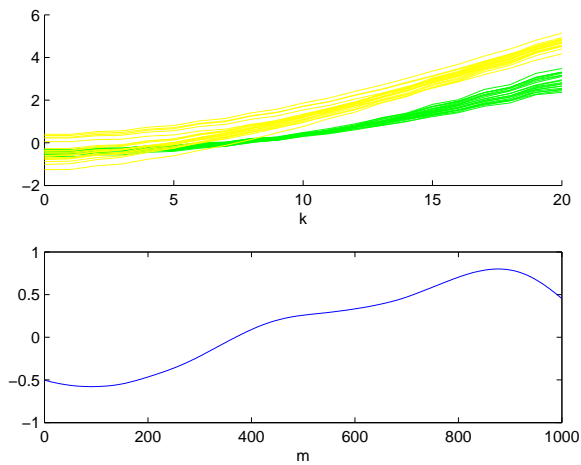}
  \caption{The top subplot shows the logarithms of the norm
of the third moment (skewness) for the $20$ instances of $P2$ (in
green) and the $20$ instances of $P3$ (in yellow). They are
plotted at each of $k=1,...,20$ iterations of (B1)-(B3) for a
specific mask that shows improved separation among these two
classes. At $k=0$ we plot the logarithms of the norm of the
skewness of the input images themselves. The bottom subplot shows
the corresponding mask defined for indexes $m=1,...,1000$.}
\end{figure}

In Figure 3 we show the third moment (skewness) for the $20$
instances of $P2$ (in green) and the $20$ instances of $P3$ (in
yellow), as we move along the dissipation process for one specific
mask that shows improved separation among these two classes. At
$k=0$ we pictured the moment distribution of the images when no
mask and no dissipation is applied. Note that there is no good
separation between these classes for $k=0,...,6$, then finally we
start to see a divergence of the clusters of moments and for $k>8$
we observe significant separation of the classes, the bottom
subplot shows the shape of the corresponding mask. This distinct
improvement of separation with tightly clustered classes happens,
with similar dynamics, for 22 out of the 100 masks that we
generated, while for the remaining masks we have no improvement in
separation as shown in Figure 4 or we have poorly clustered
classes. Interestingly, most (but not all) of the masks that show
improved separation between $P2$ and $P3$ assume negative values
for for the first few hundred largest coefficients. This generic
shape would have been difficult to predict, but is simple enough
that it raises the hope that spanning a small space of masks will
be enough in general classification problems, greatly reducing the
training computational cost, but it is likely that thousands of
masks may be necessary for very difficult classification problems.
Note that, for different pairs of classes, the masks that improve
separation may have very different shapes, for example, separation
between $P1$ and $P2$ is particularly improved by some masks that
assume negative values for the middle range coefficients and the
total number of masks for which we have improved separation with
tight clusters is smaller in this case, 13 masks out of 100. We
will use now this last observation to lower even further the
classification error rates.
\begin{figure}
  \includegraphics[ angle=0 ,width=0.7\textwidth]{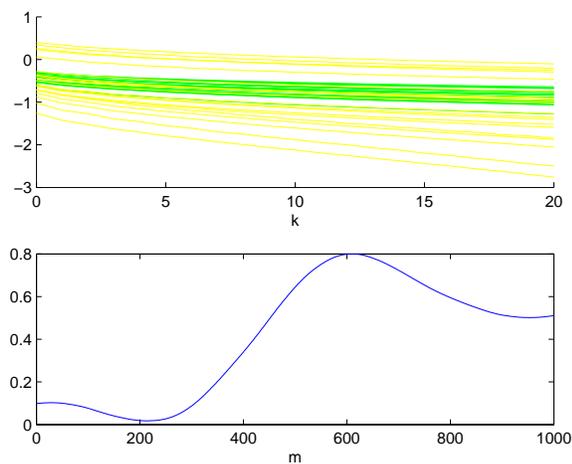}
  \caption{The top subplot shows the logarithms of the norm
of the third moment (skewness) for the $20$ instances of $P2$ (in
green) and the $20$ instances of $P3$ (in yellow). They are
plotted at each of $k=0,...,20$ iterations of (B1)-(B3) for a
specific mask that does not show improved separation among these
two classes. At $k=0$ we plot logarithms of norm of skewness of
the input images themselves. The bottom subplot shows the
corresponding mask defined for indexes $m=1,...,1000$.}
\end{figure}

The classification with $N=100$ masks and $K=20$ iterations
significantly improved error rates, but we still had a roughly
$5\%$ error rate for $P1$ and $P2$. If we relax the artificial
restriction that only $2$ best features are extracted we can
greatly improve on this result. More specifically suppose we
extract one best feature for each pair of classes, and let
$m_{ab}$ be such feature (we have a total of $6$ such features for
a 4 classes problem). The best features for each pair of classes
are shown in Figure 5. Within the setting of the previous
classification schemes, apply to each testing image a 3-NN
classification for feature $m_{ab}$ {\it using only classes $a$
and $b$} and repeat the classification for each of the $6$
features. Finally use a majority vote among all $6$
classifications to assign a class to each testing image. With the
same $N=100$ masks and $K=20$ iterations as in the previous cases
the estimated error rate are now: for $P1$ $<10^{-4}$; for $P2$
$<10^{-4}$ for $P3$ $0.008$ and for $Water$ $0.002$. If we take
only $K=1$ the errors are instead large:for $P1$ $0.027$; for $P2$
$0.050$ for $P3$ $0.246$ and for $Water$ $0.054$. With both large
number of masks and dissipation turned on {\it there is virtually
no misclassification}. This raises the issue of whether we are
overfitting the data, so as a final test of the method, we mixed
randomly all images from all classes, we divided them in four sets
of 20 images each and then we applied to these four pseudo-classes
the $6$ features classification that we just described. In this
case {\it no} mask among those that we generated and no
dissipation process is able to improve classification results,
which are consistently very poor exactly as we would expect. We
summarize our results for the $5$ classification schemes that we
used in Table 1.
\begin{table}
\begin{tabular}{|r|r|r|r|r|r|} \hline
{\it Class} &{\it K}=0 {\it p}=2&{\it K}=1 {\it p}=2&{\it K}=20 {\it p}=2&{\it K}=1 {\it p}=6&{\it K}=20 {\it p}=6\\
\hline
$Water$& $0.099$&  $0.177$&    $<10^{-4}$&  $0.054$&   $0.002$\\
$P1$   & $0.062$&  $0.050$&     $0.051$&       $0.027$&    $<10^{-4}$\\
$P2$   & $0.050$&  $0.050$&     $0.059$&       $0.050$&    $<10^{-4}$\\
$P3$&   $0.227$       &$0.481$&    $0.019$&       $0.246$&    $0.008$\\
\hline
\end{tabular}
\caption{Estimated error rates for each tested class for the $5$
versions of the 3-NN algorithm we tested. In the three cases on
the left of the table we use the $2$ best features obtained by the
functional dissipation with: $K=0$ (no dissipation), $K=1$ (one
step dissipation) and $K=20$; the right two cases use $K=1$ and
$K=20$ iterations and we select the $6$ best features such that
each optimizes separation among training sets of two different
classes. In all cases (except for $K=0$, where no mask is needed)
we used the same $N=100$ masks.}
\end{table}
\begin{figure}
  \includegraphics[ angle=0 ,width=0.7\textwidth]{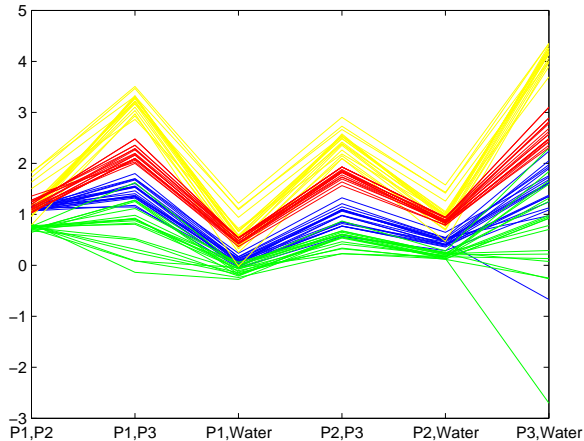}
  \caption{In this figure we plot the logarithms of the norm
of the $6$ best features among all those generated by (B1)-(B3)
with $N=100$ masks and $K=20$ iterations. Each feature is chosen
to optimize pairwise separation among training sets of two
different classes, labelled on the X-axis. Instances of of $P1$
are in blue, $P2$ in green, $P3$ in yellow and $Water$ in red.}
\end{figure}

The use of many masks allow to look at data from many different
(albeit unstructured) view points: in line with microarray
approach, we suggestively call each of the elements of the feature
vector {\it a dissipative gene}. When we display the resulting
dissipative genes in several columns, each column representing the
dissipative genes for one instance of a protein or $Water$, we
have what we can properly call a {\it functional dissipation
microarray}. In is interesting to note that, supposedly, one of
the weaknesses of matching pursuit is its inability, as a greedy
algorithm, to find an optimal representation for a given signal;
the use of randomization and dissipation turns this weakness in a
strength, at least in the setting of classification problems. This
change of perspective is in line with the idea that greedy
matching pursuit methods have a greater potential than simply
being approximations to optimal solutions. The dynamical
interaction of masks and dissipative iterations makes the signals
`flow' in the feature space and it is this flow that often carries
the most significant information on the initial conditions and
therefore on the signals themselves.

\section*{Acknowledgments}
The authors gratefully acknowledge support from DOE grant, DE-F
C52-04NA25455. We also thank the anonymous referee for very useful
and constructive remarks.

\section*{ References}
\begin{description}

\item[[1]] P. Baldi, G. W. Hatfield, W. G. Hatfield, {\it DNA
Microarrays and Gene Expression : From Experiments to Data
Analysis and Modeling}. Cambridge University Press (2002).

\item[[2]] S. Mallat, {\it A Wavelet Tour of Signal Processing},
Academic Press (1998).

\item[[3]] G. Davis, S. Mallat and M. Avelaneda, Adaptive Greedy
Approximations, Jour. of Constructive Approximation, 13 1997 (1),
pp. $57-98$.

\item[[4]] E. J. Candes and J. Romberg, Practical Signal Recovery
from Random Projections, Wavelet Applications in Signal and Image
Processing XI, Proc. SPIE Conf. vol. 5914 (2005), 59140S.

\item[[5]] E. J. Candes, J. Romberg and T. Tao, Robust uncertainty
principles: exact signal reconstruction from highly incomplete
frequency information. IEEE Trans. Inform. Theory, 52 (2004)
489-509.

\item[[6]]Y. Freund, R. Schapire, A short introduction to
boosting, {\it J. Japan. Soc. for Artifcial Intelligence}, vol. 14
n. 5 (1999), pp. $771-780$.

\item[[7]] L. Breiman. Bagging predictors. Machine Learning, 24
(1996), pp. $123-140$.

\item[[8]] P. Vincent, Y. Bengio, Kernel matching pursuit, Mach.
Learn. J. 48 (2002) (1), pp. $165-187$.

\item[[9]] J. Friedman, T. Hastie, R. Tibshirani, Additive
Logistic Regression : a Statistical View of Boosting, Annals of
Statistics, 28 (2000), pp. $337-374$.

\item[[10]] J. Friedman, Greedy function approximation: A gradient
boosting machine. Ann. Statist.  29 (2001) (5), pp. $1189-1232$.

\item[[11]] T. Hastie, R. Tibshirani, J. Friedman, {\it The
elements of Statistical Learning}, Springer (2001).

\item[[12]] K. Fukunaga. Introduction to Statistical Pattern
Recognition (2nd Edition ed.), Academic Press, New York (1990).

\item[[13]] A. Laine and J. Fan, Texture classification by wavelet
packet signatures. IEEE Trans. Pattern Anal. Mach. Intell. 15 11
(1993), pp. $1186-1191$.

\item[[14]] Q. Jin and R. Dai, Wavelet invariants based on moment
representation, Pattern Recognition Artif. Intell. 8 (1995) (3),
pp. $179–-187$.

\item[[15]] S. Noh, K. Bae, Y. Park, J. Kim, A Novel Method to
Extract Features for Iris Recognition System. AVBPA 2003, LNCS
2688 (2003), pp. $862-868$.

\item[[16]] R.W. Brockett, Dynamical systems that sort lists,
diagonalise matrices, and solve linear programming problems. Lin.
Algebra Appl. 146 (1991), pp. $79-–91$.

\item[[17]] V. N. Morozov, N. N. Vsevolodov, A. Elliott, C.Bailey,
Recognition of Proteins by Crystallization Patterns in an Array of
Reporter Solution Microdroplets, Anal. Chem. vol. 78 (2006), pp.
$258-264$.

\end{description}

\end{document}